# Ti alloyed α-Ga$_2$O$_3$: route towards wide band gap engineering

A. Barthel,[1, a)] J.W. Roberts,[2] M. Napari,[1, 3] T.N. Huq,[1] A. Kovács,[4] R.A. Oliver,[1] P.R. Chalker,[2] T. Sajavaara,[5] and F.C-P. Massabuau[1, 6, b)]

[1] *Department of Materials Science and Metallurgy, University of Cambridge, Cambridge CB3 0FS, UK*

[2] *School of Engineering, The University of Liverpool, Liverpool L69 3GH, UK*

[3] *Zepler Institute for Photonics and Nanoelectronics, University of Southampton, Southampton SO17 1BJ, UK*

[4] *Ernst Ruska-Centre for Microscopy and Spectroscopy with Electrons and Peter Grünberg Institute, Forschungszentrum Jülich, 52425 Jülich, Germany*

[5] *Department of Physics, University of Jyväskylä, FI-40014 Jyväskylä, Finland*

[6] *Department of Physics, SUPA, University of Strathclyde, Glasgow G4 0NG, UK*

a) ab2301@cam.ac.uk

b) f.massabuau@strath.ac.uk

The suitability of Ti as a band gap modifier for α-Ga$_2$O$_3$ was investigated, taking advantage of the isostructural α-phases and high band gap difference between Ti$_2$O$_3$ and Ga$_2$O$_3$. Films of Ti:Ga$_2$O$_3$, with a range of Ti concentrations, synthesized by atomic layer deposition on sapphire substrates, were characterized to determine how crystallinity and band gap vary with composition for this alloy. The deposition of crystalline α-(Ti$_x$Ga$_{1-x}$)$_2$O$_3$ films with up to x~5.3%, was demonstrated. At greater Ti concentration, the films became amorphous. Modification of the band gap over a range of ~270meV was achieved across the crystalline films and a maximum change in band gap from pure α-Ga$_2$O$_3$ of ~1.1 eV was observed for the films of greatest Ti fraction (61% Ti relative to Ga). The ability to maintain a crystalline phase at low fractions of Ti, accompanied by a significant modification in band gap shows promise for band gap engineering and the enhancement in versatility of application of α-Ga$_2$O$_3$ in optoelectronic devices.

# I. INTRODUCTION

Alpha phase gallium oxide (α-Ga$_2$O$_3$) is an ultra-wide band gap semiconductor, with most measurements of its band gap lying between 5.1 eV and 5.3 eV [1-5]. It is of particular interest for application in solar-blind ultraviolet (UV) photodetectors [2,4,6,7]. Uses for photodetectors that can absorb efficiently in this regime, of wavelengths <285 nm, include water and air purification systems [8], flame detection, UV astronomy, missile defence systems and engine monitoring [4,9].

α-Ga$_2$O$_3$ is a metastable phase of Ga$_2$O$_3$, a polymorphic group III sesquioxide, with commonly reported phases α, β, γ, δ, and ε [10] as well as the more recent κ [11]. Previous research on this material has been mostly focused on the stable, monoclinic β phase [10,12,13], however, due to recent advances in thin film growth techniques, such as mist chemical vapour deposition (mist-CVD) [1,14,15] and atomic layer deposition (ALD) [4,5,16], it has become possible to synthesise high quality films of α-Ga$_2$O$_3$. These films are grown epitaxially on sapphire (α-Al$_2$O$_3$), which shares its rhombohedral corundum crystal structure1 (inset Figure 1) with α-Ga$_2$O$_3$. This has been achieved at temperatures as low as 250 °C by ALD [4,5,16] and the material has been successfully integrated into solar-blind UV photodetectors, already showing an advantage over photodetectors based on α-Ga$_2$O$_3$, by having shorter response times [4].

Apart from sapphire, the corundum crystal structure is also shared by many other semiconducting sesquioxides [17-19], as shown in Figure 1, providing great potential for band gap engineering [15,20]. Previously, alloying of corundum phase Ga$_2$O$_3$ with Al$_2$O$_3$ [21], In$_2$O$_3$ [22], Cr$_2$O$_3$ [18], Fe$_2$O$_3$ [18,19] and Rh$_2$O$_3$ [23] has been attempted.

The aim of this work is to study the feasibility of using Ti as an effective band gap modifier for α-Ga$_2$O$_3$, by characterizing a number of oxide films grown by ALD with different Ti to Ga ratios. α-Ti$_2$O$_3$ adopts the corundum crystal structure [36,40] with lattice parameters: a = 5.157 Å and c = 13.613 Å [36], giving a relatively small lattice mismatch of about 3.5% with α-Ga$_2$O$_3$ (a = 4.983 Å and c = 13.433 Å [33]). Its direct band gap of 0.1 eV [26,27] is very small relative to that of α-Ga$_2$O$_3$, such that a wide range of band gaps may be achievable, provided that an alloy of the two sesquioxides exhibits miscibility and crystallinity across a range of Ti:Ga ratios. This may be inhibited if Ti adopts a +4 oxidation state and preferentially forms TiO$_2$, which would not have a corundum structure. Another property of α-Ti$_2$O$_3$ that is of interest is that it is a p-type semiconductor [27]. Alloying with Ti$_2$O$_3$ could thus provide a route to achieve p-type conductivity in α-Ga$_2$O$_3$, as was also demonstrated in an α phase Rh:Ga$_2$O$_3$ alloy by Kaneko *et al.* [23].

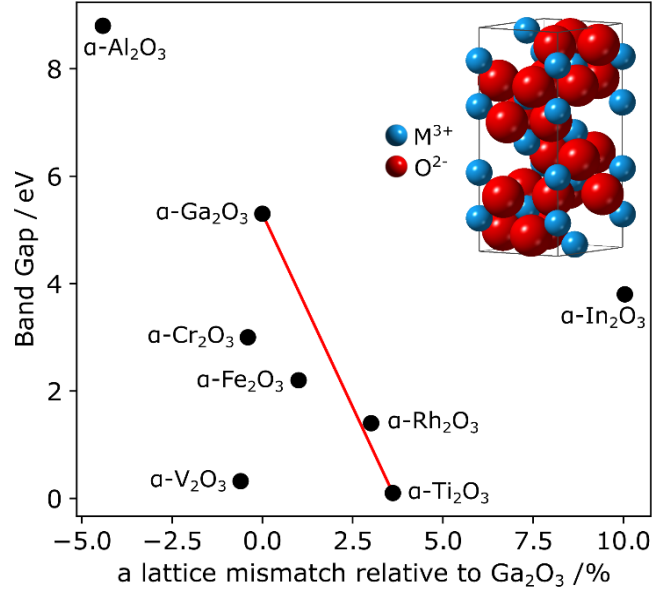

**Figure 1.** Diagram of the corundum phase semiconducting sesquioxide design space, centered on α-Ga$_2$O$_3$. The band gaps of the materials [1,24-32] are plotted against their a-lattice parameters relative to α-Ga$_2$O$_3$ [33-39]. The red line indicates the alloys of α-Ti$_2$O$_3$ and α-Ga$_2$O$_3$, assuming that the band gap varies linearly with the lattice constants. Inset: Rhombohedral, corundum crystal structure of M$_2$O$_3$ (M: metal).

## II. EXPERIMENTAL METHODS

Ga$_2$O$_3$ and Ti:Ga$_2$O$_3$ thin films were grown using an Oxford Instruments OpAL Plasma Enhanced Atomic Layer Deposition (PEALD) Reactor. All films were grown on 0.25° miscut c-plane sapphire substrates with a temperature of 250 °C and chamber wall temperatures set to 150 °C. Triethylgallium (TEGa) from Epichem and Titanium(IV) Isopropoxide (TTIP) from Sigma Aldrich were used as Ga and Ti precursors, respectively. The TEGa and TTIP precursors were held at 30 °C and 80 °C, respectively, with line temperatures for both precursors set at 90 °C and 100 °C, increasing in temperature closer to the reaction chamber. One cycle of Ga$_2$O$_3$ consisted of 0.1 s TEGa with 100 sccm Ar bubbling, 5 s 100 sccm Ar purge, 3 s O2 ow stabilisation, 5 s 20 sccm 300W O$_2$ plasma, 5 s 100 sccm Ar purge. One cycle of TiO$_x$ consisted of 2 s TTIP with 100 sccm Ar bubbling, 10 s 100 sccm Ar purge, 0.04 s H$_2$O, 10 s 100 sccm Ar purge. The chosen growth parameters were adapted from Ref.[41]. Supercycles of Ga$_2$O$_3$ and TiO$_x$ were used to produce Ti doped Ga$_2$O$_3$ films, with cycle ratios (TiO$_x$: Ga$_2$O$_3$) of 0:1, 1:32, 1:19, 1:9, 1:4 and 1:1, which are hereafter referred to as samples 0%Ti, 3%Ti, 5%Ti, 10%Ti, 20%Ti and 50%Ti, respectively. 500 total cycles were used for the 0%Ti film, 429 cycles for the 3 %Ti film and 400 cycles for the remaining films. Film thicknesses after growth were measured using a Horiba-Yvon spectroscopic ellipsometer fitted to a mixed Cauchy α- Ga$_2$O$_3$/TiO$_2$ model and are shown in Table I.

The Ti:Ga ratio of the samples was determined using Rutherford backscattering spectrometry (RBS) with 1.615MeV He$^+$ incident beam from a 1.7 MV Pelletron accelerator. The samples were tilted to 5° and the scattering angle was 165°. The measured spectra were analysed using SimNRA program.

The crystallinity of the samples was assessed by X-ray diffraction (XRD). A PANalytical Empyrean diffractometer was used with a Cu source and a hybrid two bounce primary monochromator giving Cu Kα1 radiation and either a two-bounce Ge crystal analyser, for 2θ-ω, or a PIXcel detector, for reciprocal space maps (RSMs).

High-angle annular dark field scanning transmission electron microscopy (HAADF-STEM) using an aberration-corrected FEI Titan [42] operated at 200 kV was used to observe the sub-surface structure of the samples observed in cross-section. The annular dark field detector semi-angle used was 69.1 mrad. Compositional mapping was obtained using energy dispersive X-ray spectroscopy (EDX) in the same microscope and strain mapping was obtained using geometrical phase analysis43. The samples were prepared for imaging using standard mechanical grinding followed by Ar$^+$ ion milling at 5 kV and cleaning at 0.1-1 kV.

The surface morphology of the samples was investigated by atomic force microscopy (AFM) in a Bruker Dimension Icon operated in peak force tapping mode. SCANASYSTAIR tips with a nominal radius of 2nm were used.

The band gaps of the films were determined from transmittance spectra of the films, measured using a Cary 7000 UV-VIS-NIR spectrometer in the range 200-800 nm. The system was calibrated for 100% and 0% transmittance.

| Sample | Growth Ti:Ga ratio | Film thickness /nm | Composition /% | RMS roughness /nm | Band Gap /eV |
|---|---|---|---|---|---|
| 0%Ti | 0:1 | 33 ± 3 | 0 | 0.71 ± 0.01 | ~5.04 |
| 3%Ti | 1:32 | 22 ± 2 | 3.7 ± 0.4 | 0.49 ± 0.02 | ~4.88 |
| 5%Ti | 1:19 | 21 ± 2 | 5.3 ± 0.4 | 0.30 ± 0.01 | ~4.77 |
| 10%Ti | 1:9 | 21 ± 2 | 12.8 ± 0.8 | 0.22 ± 0.01 | ~4.78 |
| 20%Ti | 1:4 | 21 ± 2 | 23.1 ± 1.0 | 0.19 ± 0.01 | ~4.63 |
| 50%Ti | 1:1 | 16 ± 2 | 61 ± 3 | 0.16 ± 0.01 | ~3.91 |

**Table 1.** Summary of film characteristics. Samples are named after the percentage of Ti ALD cycles used for growth. Film thicknesses, obtained by ellipsometry, are estimates as an appropriate fitting model for α-Ga$_2$O$_3$, amorphous Ga$_2$O$_3$ or α-Ti$_2$O$_3$ was unavailable. Composition, measured by RBS, represents the amount of Ti relative to Ga in the films. RMS roughness and band gaps were determined by AFM and UV-vis, respectively.

**III. RESULTS AND DISCUSSION**

RBS was used for compositional analysis of the films. The main quantity of interest was the ratio of Ti to Ga in the films. RBS spectra were obtained for each film, showing peaks for Ga and Ti (except in the pure Ga$_2$O$_3$ sample) as well as steps associated with Al and O from the sapphire substrate. The compositions x, here defined as x=at.%Ti /(at.%Ti+at.%Ga), attained from the integrated counts of the Ti and Ga peaks in the RBS spectra and corrected for by the respective Rutherford scattering cross sections, are tabulated in Table I. The film compositions determined by RBS, are in relatively good agreement with those predicted from the ALD cycle ratios, deviating by about 10-30%. Deviations resulting from unequal growth rates of the different species are expected in ALD, with the growth being strongly affected by several factors, including temperature, chemistry of the precursors used, and the growth surface [44,45].

XRD was used to study the crystallinity of the films. Prior work [16] showed that α-$Ga_2O_3$ grown by ALD on c-plane sapphire had the same crystal orientation as the substrate. However, rather than a homogeneous film, it was reported that the film consisted of α-$Ga_2O_3$ columns separated by amorphous material. The presence of crystalline corundum phase material in the Ti:$Ga_2O_3$ films was validated by measurement of high intensity, symmetric 0006 reflections. The absence of other characteristic corundum phase reflections, as in the prior work [16], suggests the addition of Ti does not alter the crystal orientation of the Ti:$Ga_2O_3$ film relative to the substrate.

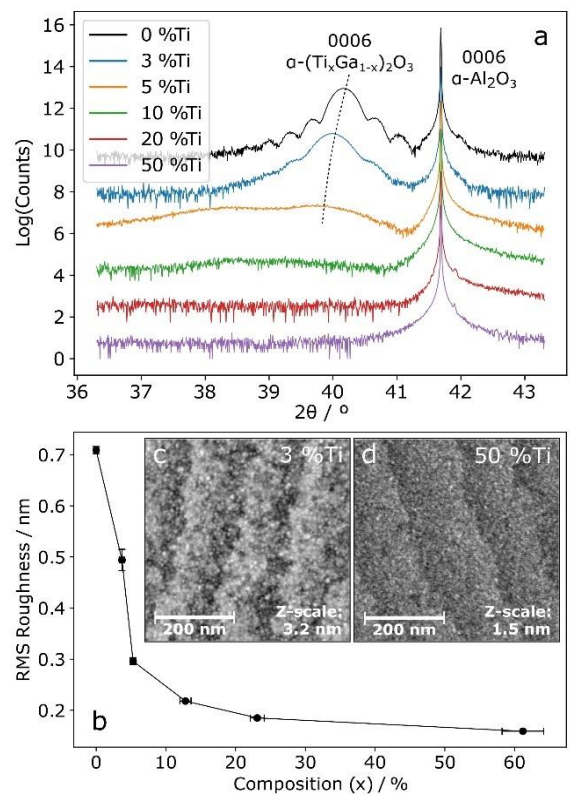

**Figure 2.** (a) 2θ-ω XRD scans of the symmetric 0006 reflection of the α phase. (b) Variation of RMS surface roughness with Ti concentration. (c, d) 500nm size AFM images of the 3%Ti (c) and 50%Ti (d) samples.

The measured 2θ-ω XRD scans (Figure 2(a)) show the 0006 peak from the sapphire substrate and, for the 3 samples of lowest Ti fraction, also a peak from the film. The pure α-$Ga_2O_3$ sample (sample 0%Ti) gives the peak of greatest intensity at 2θ = 40.16°, which is close to literature values [16,33]. Differences in lattice parameter may be due to residual strain from the epitaxial relationship with the substrate. (Annealing the samples at *ca.* 400 °C has been shown to release that residual strain [7].) Interference fringes can be distinguished on the base of the diffraction peak, with a spacing that is representative of the film thickness – here 28 nm, in reasonable agreement with the thickness estimated by ellipsometry (Table I).

As the concentration of Ti increases, the intensity of the peaks decreases rapidly, vanishing for sample 10%Ti, suggesting that the samples are undergoing a phase change, most likely to an amorphous phase. There is a second peak in the scan for the 5%Ti sample, which appears at a lower angle (~38°), possibly indicating the emergence of one or more other phase, such as β-$Ga_2O_3$ and ε-$Ga_2O_3$ [7,46]. However, the peak is less intense than the already weak α-$Ga_2O_3$ 0006 reflection, suggesting that at this concentration of Ti, the alloy is already mostly amorphous.

The α-Ti:$Ga_2O_3$ peak also shifts to smaller angles as the Ti content increases, indicating that the lattice parameter of the crystalline phase is increasing. This is expected, as the lattice parameters of $Ti_2O_3$ are larger and suggests that Ti has been incorporated into the films to form a $(Ti_xGa_{1-x})_2O_3$ alloy. However, the measured c lattice parameters exceed those expected from the compositions calculated from RBS data, using literature values for lattice parameters and assuming the applicability of Vegard's Law [47,48]. A likely cause of this is that the $(Ti_xGa_{1-x})_2O_3$ layers are compressively strained onto the sapphire substrate. However, due to insufficiently strong signals, it was not possible to quantify the strain in the films by conducting a strain analysis from RSMs of symmetric and asymmetric reflections.

The surface topography of each sample was obtained by AFM and root mean square (RMS) roughness determined at a 500 nm scan size. The AFM images (Figure 2(c-d)) are representative of the topographies observed for all samples. The surface was smooth on a nanometer scale, as shown by the RMS roughness, given in Table I and Figure 2(b).

The surface of all samples also bears ledges with a constant spacing, mostly independent of Ti content. These ledges likely arise from the morphology of the sapphire substrates, which, due to having a miscut angle of 0.25°, also have steps on the surface. These provide two distinct regions: a flat (0001) oriented surface and the step. Given preferential growth at one of these sites in ALD, a step pattern with the same spacing would be expected. A topographical scan of a pristine sapphire wafer (not shown here) shows similar step widths as Figure 2(c-d).

The surfaces of the low Ti fraction samples (Figure 2(c)) exhibit small grains of the order of 1-10 nm, presumably due to the termination of α phase columns [5,16]. As the Ti fraction increases (Figure 2(d)) these features cannot be identified anymore, which may indicate that the films are amorphous, in line with our XRD results. The plot of the RMS roughness vs Ti fraction (Figure 2(b)) shows that the roughness decreases quickly with Ti fraction. This decrease in RMS roughness coincides with the disappearance of the small grain features as the Ti fraction increases. We note however that the magnitude of the decrease could be amplified by the reduced film thickness of the Ti-containing samples. This trend (together with the images Figure 2(c-d) as well as XRD results Figure 2(a)) implies that the low Ti fraction (<10%) films are crystalline, whilst the films with high Ti fraction (>10%) are amorphous.

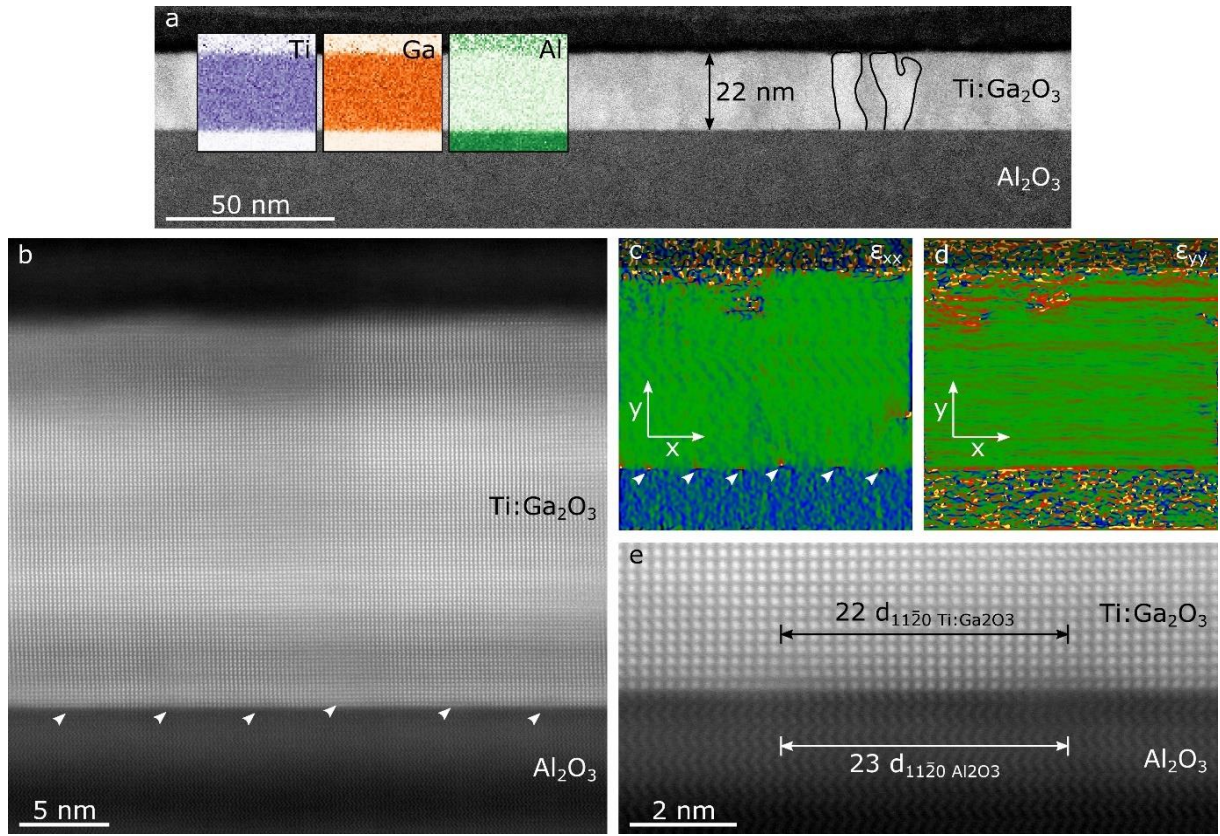

**Figure 3.** (a-b) HAADF-STEM images of a cross-section of sample 3%Ti, showing the Ti:Ga$_2$O$_3$ film and its interface with the sapphire substrate. EDS maps are overlaid in (a), showing a uniform distribution of Ti and Ga throughout the film. Two columnar regions of crystallinity in the Ti:Ga$_2$O$_3$ film are also outlined in (a). Strain in the film, perpendicular (c) and parallel (d) to growth direction, was obtained from geometrical phase analysis. Regularly spaced misfit dislocations are indicated by arrows in (b-c). (e) High resolution HAADF-STEM image of film-substrate interface, clearly showing 2 misfit dislocations and their separation

STEM was used to analyse a cross section of sample 3 %Ti, allowing the interface between substrate and film to be studied as well as giving a local insight to crystallinity, strain and composition across the film. The results are compiled in Figure 3, showing large, vaguely columnar regions of crystallinity in the film, outlined in Figure 3(a), with, in places, regions of amorphous material. This is in agreement with prior work on ALD grown, pure Ga$_2$O$_3$ [16]. The film thickness measured by HAADF-STEM images (Figure 3(a)) is consistent with that measured by ellipsometry and the film seems to have a fairly uniform thickness, as suggested by the low surface roughness from AFM. From the high resolution data (Figure 3(b)) the epitaxial relationship with the sapphire substrate was identified to be $\langle 11\bar{2}0\rangle_{Al2O3} \parallel \langle 11\bar{2}0\rangle_{(Ti_xGa_{1-x})2O3}$ and $[0002]_{Al2O3} \parallel [0002]_{(Ti_xGa_{1-x})2O3}$, in agreement with previous studies [16].

Observation of the interface region reveals the presence of a high density of periodically spaced misfit dislocations, with an average spacing of 4.9±0.4 nm. Apart from these, there are also some dislocations located within the film, *ca.* 1-3nm from the interface (not shown here) although these are more rare. The misfit dislocations at the interface can be easily seen in both Figure 3(b) and (c) (indicated by arrows), the latter showing them as dipoles of tensile and compressive strain. Figure 3(e) shows two misfit dislocations at greater magnification such that it is possible to count the number of lattice planes between them. They are separated by $23d_{11\bar{2}0\ Al2O3}$ or $22d_{11\bar{2}0\ (TixGa1-x)2O3}$. This indicates that the film is almost fully relaxed (nominally $22d_{11\bar{2}0\ Al2O3}$= 5.232 nm, $21d_{11\bar{2}0\ Ga2O3}$=5.232 nm and $21d_{11\bar{2}0\ (TixGa1-x)2O3}$=5.238 nm assuming x = 3.7%). The strain maps in Figure 3(c-d) confirm that all the strain relaxation occurs at the interface, and that the strain is otherwise uniform throughout the film. The strain relaxation of the film observed in STEM is in contradiction with the XRD data, which instead suggests that the film is strained. A possible explanation for this contradiction between XRD and STEM may be that these probe the material at different scales. TEM focuses on a small nanometre-scale region of the sample – typically looking at the scale of an individual column – whereas XRD looks at average/global properties of the film. Another possibility is that the observed imperfect spacing of the misfit dislocations for $(Ti_xGa_{1-x})_2O_3$ is unable to fully accommodate the strain.

EDS analysis of the film showed that the distribution of Ti across the film was uniform, with no visible segregation (Figure 3(a)). Compositional analysis using Cliff-Lorimer method yielded a composition of x ~ 5%, which is in fair agreement with the RBS data (we note however that EDS quantification is here less reliable than the RBS data, because the sample was imaged on the zone-axis for the α phase and was therefore susceptible to interference from electron channeling).

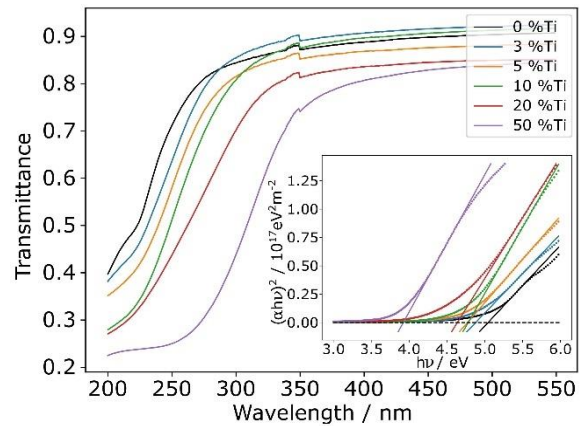

**Figure 4.** Transmittance spectra for the Ti:Ga$_2$O$_3$ samples. The inset shows Tauc plots evaluated using the transmittance data for each film. The straight, solid lines are the fits to the linear region of the Tauc plots and their intercepts with the dotted black line (α = 0) gives the band gaps of the respective films.

UV-vis of the films was conducted to attain their optical band gaps. Only transmittance spectra were taken. Ignoring reflected intensity is justified by the fact that the films were transparent and only weakly reflecting to the eye, at normal incidence. Using the assumption of negligible reflectance, Beer-Lambert law could be applied to calculate absorptivity from the transmittance [2,49]. Film transmittances were isolated from the transmittances of the samples by measuring that of a reference pristine sapphire substrate. Absorptivity could then be used to produce Tauc plots using the relation in equation (1), from which the band gap was determined. A Tauc exponent of n =1/2 for a direct band gap and allowed transition was used, as was also applied in other works [1-5]:

$$\alpha h\nu \propto (h\nu - E_g)^n$$

The collected transmittance spectra for the films are depicted in Figure 4 with an inset of the Tauc plots used to attain the band gaps. Even without the Tauc plots it is already obvious that the absorption edge shifts to longer wavelengths as Ti concentration in the films increases. Above the absorption edge, the transmittance is very high, suggesting that the assumption of little reflection or scattering, when calculating absorptivity is justified (the small discontinuity in the spectrum at ~350nm is a systematic instrumental error, due to the source change that occurs in the system at that wavelength). The transmittance at wavelengths longer than the absorption edge seems to decrease with increasing Ti concentration, which could be due to increased scattering or reflection from the films, which are not entirely crystalline.

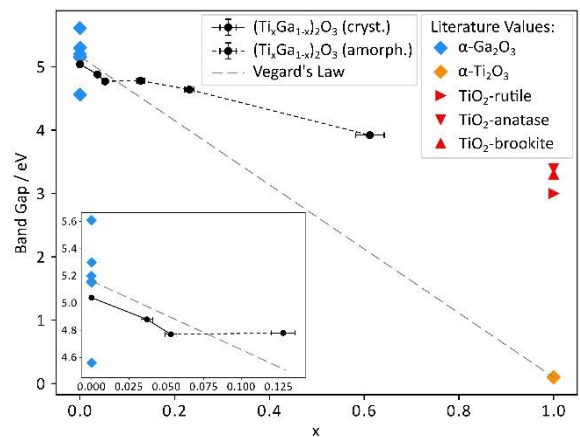

**Figure 5.** Plot showing the measured variation of band gap with the RBS composition alongside literature values of band gaps for pure oxides [1-5,27,50-56] and a Vegard's law trend between α-Ga$_2$O$_3$ and α-Ti$_2$O$_3$. The inset is an enlargement of the low Ti region. The plot also differentiates between compositions based on whether crystallinity was observed by XRD.

The decrease in band gap with increasing Ti concentration can be clearly seen in Figure 5. The figure also shows literature values of band gaps for pure α-Ga$_2$O$_3$, Ti$_2$O$_3$ and 3 common phases of TiO$_2$: rutile [53,54], anatase [55] and brookite [56]. The band gap measured for pure α-Ga$_2$O$_3$ is well within the range of the literature [1-5,50,51]. The linear trend in band gap with composition predicted by Vegard's law for α-(Ti$_x$Ga$_{1-x}$)$_2$O$_3$ has also been plotted. The measured band gaps seem to follow this trend for the samples with lowest Ti concentration, up to 5%Ti, which are also the films that were determined to contain crystalline α phase, with a uniform distribution of Ti, by XRD and TEM. Over this range of composition the band gap varies by ~270 meV, which confirms that Ti:Ga$_2$O$_3$ is a viable route to band gap tuning in the UV. Taking amorphous films into account, an even greater band gap variation of ~1.12 eV was achieved. However, beyond 5%Ti, the 3 samples of greatest Ti concentration deviate from the original trend for one that may still be linear, but with a more shallow slope. Given that only 3 data points each are available for crystalline and amorphous materials, it is not currently possible to determine if the relationship between band gap and composition follows Vegard's law or whether there is any bowing.

## IV. CONCLUSION

In conclusion, films of Ti:$Ga_2O_3$ with a range of Ti concentrations were synthesized by ALD and the resulting film crystallinity and band gap characterized. The band gap variation with composition seemed to follow two separate linear trends that coincide with different film crystallinities. We demonstrate the deposition of crystalline α-$(Ti_xGa_{1-x})_2O_3$ films with up to x ~ 5.3%, resulting in a change in band gap from pure α-$Ga_2O_3$ of up to ~ 270 meV. On the other hand, films deposited with a greater Ti concentration were amorphous, leading to change in band gap from pure α-$Ga_2O_3$ of up to ~ 1.1 eV for the films containing 61% Ti relative to Ga. This study is a promising proof-of-principle that Ti can be used to modify the bandgap of α-$Ga_2O_3$ materials and opens the path for the fabrication of wavelength-specific optoelectronic devices operating in the UV.

## V. ACKNOWLEDGMENTS

This project is funded by the Engineering and Physical Sciences Research Council (EPSRC Grants No. EP/P00945X/1 and No. EP/M010589/1). T.N.H. acknowledges funding from the EPSRC Centre for Doctoral Training in Graphene Technology (Grant No. EP/L016087/1). This project has received funding from the European Union's Horizon 2020 research and innovation programme under grant agreement No 823717 – ESTEEM3.